\documentclass[fleqn,twoside]{article}
\usepackage{espcrc2}

\usepackage{graphicx}
\usepackage{epsfig,german}
\usepackage[figuresright]{rotating}

\newcommand{\AmS}{{\protect\the\textfont2
  A\kern-.1667em\lower.5ex\hbox{M}\kern-.125emS}}

\title{Parametrizations of Quark Confinement in Production Processes}

\author{C.S.~Fischer\thanks{Talk given at the workshop {\it The Spin Structure of the Proton},
Trento, July 23rd to 28th, 2001.}\\
\vspace*{0.2cm}
{Institute for Theoretical Physics, University of T"ubingen\\ 
        Auf der Morgenstelle 14, 72070 T"ubingen, Germany}
	}
\begin{document}
\begin{abstract}
Baryons are described as bound states of quarks and scalar as well as axialvector diquarks.
In order to effectively parametrize confinement several {\it ansaetze} for the dressing
functions of the constituent propagators are used.
The corresponding results for electromagnetic and strong form factors differ only slightly.
On the other hand observables from production processes show the permissibility 
of different dressing functions. 
\vspace{1pc}
\end{abstract}

% typeset front matter (including abstract)
\maketitle

The aim of the present investigation is to set up a model describing baryon properties at intermediate 
energies, where great experimental progress has been made in the last decade. 
In order to capture essential features of QCD at least two requirements 
should be met: the model has to be Poincar\'e covariant and it has to incorporate
confinement to avoid unphysical thresholds. 
The diquark-quark model fulfills both demands.

Quark-quark correlations inside the baryon are dominated by separable contributions
that carry the quantum numbers of the scalar and the axialvector diquark \cite{Wetzorke:2000ez}.
Therefore baryons can be described by a quark-diquark Bethe-Salpeter (BS) equation, with the 
constituents interacting via quark exchange \cite{Oettel:1998bk,Alkofer:2000wg}.
We solve the BS equation without reduction (cf. \cite{Oettel:2000xx}) to obtain covariant vertex 
functions for octet and decuplet baryons. 

The central input of the model is the {\it ansatz} for the dressing function of the quark and
diquark propagators.  We 
employ three different methods to parametrize confinement: 
first we use an exponential dressing function to shift the poles of the bare propagators to
timelike infinity ('exp-propagator'), second we split the real poles into complex conjugate
pairs ('cc-propagator') and third we employ a nonanalytic dressing function that
removes the poles completely ('tanh-propagator') \cite{Ahlig:2001qu}. 

The electric 
and magnetic form factors of the proton as well as the mass spectrum of the octet baryons
serve to fix the parameters of the model. Special care is taken of electromagnetic current
conservation \cite{Oettel:2000gc}.
Although the contributions from the axialvector diquark
to the vertex functions of the nucleon are at the 10\% level, it turns out that they are vital
to reproduce the magnetic moments \cite{Oettel:2000jj}. A summary of the
electromagnetic properties of the nucleon in our model can be found in \cite{Alkofer:2001ne}. 

Having fixed the parameters, the results for the strong form factors are  
predictions. We obtain a soft pion nucleon form factor $g_{\pi NN}$ and
good agreement with the experimental value 
at zero momentum. 
Due to flavour symmetry breaking the results for the kaon nucleon lambda form factor $g_{KN\Lambda}(p^2)$
are smaller than the one for $g_{\pi NN}(p^2)$ for all momenta. 
In general all three {\it ansaetze} for the dressing functions of the propagators lead to 
similar results for form factors \cite{Ahlig:2001qu}.

This is different for production processes. In the diquark spectator approximation there are two 
types of diagrams that describe kaon photoproduction, $\gamma p \rightarrow \Lambda K$.
In the s-channel diagram both, photon and kaon, couple to a single quark line, whereas in the
t-channel the photon couples directly to the kaon.
\begin{figure}
\epsfig{file=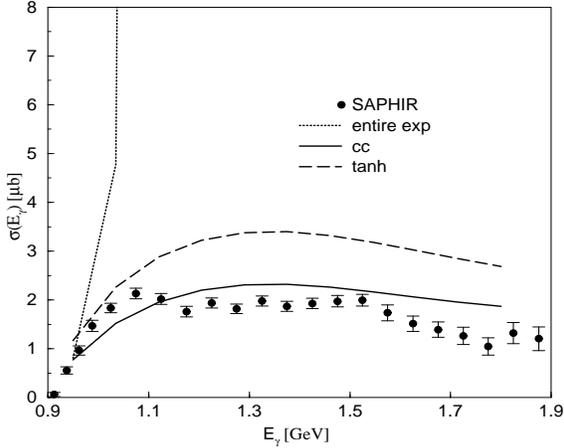,width=7.5cm,height=6.8cm}
\vspace*{-1.3cm}
\caption{Total cross section of $\gamma p \rightarrow \Lambda K$ 
compared to the data of the 
SAPHIR-collaboration \cite{Tran:1998qw}.}
\label{cross}
\vspace*{-0.4cm}
\end{figure} 
Fig.~\ref{cross} shows that the parametrization with the exponential dressing function overshoots the 
data by orders of magnitude. This is because one of the propagators in the s-channel diagram is
probed for large timelike momenta where the exponential dressing blows up very quickly.
We therefore consider the exponential dressing {\it ansatz} to be ruled out for the whole class of models
that resolve baryons 
into its quark substructure \cite{Ahlig:2001qu}.
The other two parametrizations describe the data reasonably well. Here the t-channel diagram dominates.
It has been shown, however, that interference between the two channels even leads to a good
description of the substructure of the data \cite{Ahlig:2000xm}.  

For associated strangeness production, $pp \rightarrow pK\Lambda$, we obtain a similar picture
with respect to the dressing functions. Again the exponential dressing
overshoots the data massivly, whereas the cc-propagator and the tanh-propagator give
cross sections with the right order of magnitude. The spin depolarisation tensor $D_{NN}$ 
provides a measure for the relative strengh of diagrams with pion and kaon exchange 
between the two protons. Pion exchange leads to positive values of $D_{NN}$ whereas
\begin{figure}
\epsfig{file=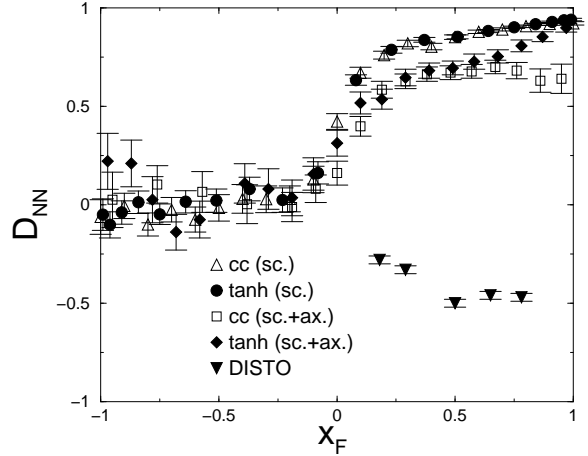,width=7.5cm}
\vspace*{-1.5cm}
\caption{Depolarisation tensor $D_{NN}$ for $pp \rightarrow pK\Lambda$ 
compared to the data of the 
DISTO-collaboration \cite{Balestra:1999br}.}
\label{dnn}
\vspace*{-0.4cm}
\end{figure} 
kaon exchange gives negative contributions. Fig.~\ref{dnn} shows, that the experimental data seem to favour kaon exchange, 
whereas pion exchange diagrams dominate in our model \cite{Ahlig:2001qu}. We regard this as a sign for missing 
contributions in the kaon exchange channel.  

{\bf Acknowledgements}

I thank S.~D.~Bass, A.~De.~Roeck and A.~Deshpande for organising this very interesting workshop
and for the invitation. I am grateful to S.~Ahlig, R.~Alkofer, M.~Oettel, H.~Reinhardt and 
H.~Weigel for their collaboration on the work presented here. Thanks for the support
by the European Graduate School Basel-T"ubingen and by COSY under contract no. 4137660.

\end{document}